\newcommand{\stdpackages}{
  \usepackage{amsmath}
  \usepackage{amssymb}
  \usepackage{amsfonts}
  \allowdisplaybreaks
  \usepackage{amsthm}
  \usepackage{eucal}
  \usepackage[final]{graphicx}
  \usepackage{color}
  \usepackage{psfrag}
  \usepackage{fancyhdr}
  \renewcommand{\headrulewidth}{.5pt}\renewcommand{\footrulewidth}{.0pt}\cfoot{}
  \setlength{\headsep}{10mm}
  \fancyhead[OL]{\it\theauthor---\today}
  \fancyhead[ER]{\leftmark}
  \fancyhead[OR,EL]{\thepage}
  \fancyfoot[EL,OR]{}

  \newcommand{\draft}{\usepackage[light,first]{draftcopy}\draftcopyName{draft}{350}}
  \newcommand{\labels}{\usepackage{blabla/tex/showlabels}}
  \newcommand{\maple}{\usepackage{maple2e}}
  \newcommand{\makeidx}{\usepackage{makeidx}\makeindex}
  \newcommand{\chicago}{\usepackage{blabla/tex/chicago}\bibliographystyle{blabla/tex/chicago}\renewcommand{\refname}{References\thispagestyle{empty}\renewcommand{\refname}{}}}
  \newcommand{\numberlines}{
    \usepackage[]{blabla/tex/lineno} 
    \linenumbers
    \modulolinenumbers[5]
  }
  \newcommand{\pdflatex}{
    \definecolor{bluecol}{rgb}{0,0,0}
    \definecolor{greencol}{rgb}{0,.6,0}
    \usepackage[
    pdftex,
    bookmarks,
    bookmarksnumbered,
    colorlinks,
    urlcolor=bluecol,
    citecolor=bluecol,
    linkcolor=bluecol,
    pagecolor=bluecol,
    pdfborder={0 0 0},
    pdfpagemode=None, 
    pdfauthor={Marc Toussaint}
    ]{hyperref}
    \DeclareGraphicsExtensions{.jpg,.pdf}
    \renewcommand{\r}{\varrho}
    \renewcommand{\l}{\lambda}
    \renewcommand{\L}{\Lambda}
    \renewcommand{\s}{\sigma}
    \renewcommand{\O}{\Omega}
    \renewcommand{\SS}{{\cal S}}
    \renewcommand{\boldsymbol}{}
  }
}
\newcommand{\stdtheorems}{
  \theoremstyle{plain}

  \theoremstyle{definition}
  
  \theoremstyle{remark}

}
\newcommand{\stdstyle}[1]{
  \stdpackages
  \stdtheorems
  \usepackage{mt}
\renewcommand{\labelenumi}{\textbf{(\roman{enumi})}}
  \newcommand{\blockindent}{3ex}
  \renewcommand{\baselinestretch}{#1}
  \renewcommand{\arraystretch}{1.2}
  \renewcommand{\topfraction}{1}
  \renewcommand{\bottomfraction}{1}
  \renewcommand{\textfraction}{0}
  \columnsep 5ex
  \parindent 3ex
  \parskip 1ex

\parindent 0pt
\topsep 4pt plus 1pt minus 2pt
\partopsep 1pt plus 0.5pt minus 0.5pt
\itemsep 2pt plus 1pt minus 0.5pt
\parsep 2pt plus 1pt minus 0.5pt
\parskip .5pc

  \setcounter{tocdepth}{3}
  \setcounter{secnumdepth}{2}

  \usepackage{./geometry}
  \geometry{a4paper,dvips,hdivide={35mm,*,35mm},vdivide={25mm,*,35mm},twosideshift=0mm}

  \thispagestyle{fancy}
  \pagestyle{fancy}
  \renewcommand{\maketitle}{\mytitle}
  \renewenvironment{abstract}
  {\begin{rblock}\hrule\vspace{2ex}{\bf Abstract.~}\small}
    {\vspace{3ex}\hrule\end{rblock}\vspace{5ex}}
  \usepackage{mt}
}
\newcommand{\cleardefs}{
  \renewcommand{\article}[2]{}
  \renewcommand{\book}[2]{}
  \renewcommand{\draft}{}
  \renewcommand{\labels}{}
  \renewcommand{\maple}{}
  \renewcommand{\makeidx}{}
  \renewcommand{\chicago}{}
  \renewcommand{\pdflatex}{}
  \renewcommand{\header}{}
}

\newcommand{\article}[2]{
  \documentclass[#1pt,twoside,fleqn]{article}
  \stdstyle{#2}
  \macros
}
\newcommand{\nips}{
  \documentclass{article}
  \usepackage{blabla/tex/nips2003e,times}
  \stdpackages\macros
}
\newcommand{\ijcnn}{
  \documentclass[10pt,twocolumn]{blabla/tex/ijcnn}
  \stdpackages\macros
  \bibliographystyle{abbrv} 
}
\newcommand{\springer}{
  \documentclass{blabla/tex/springer_llncs}
  \renewcommand{\theenumi}{\alph{enumi}}
  \renewcommand{\labelenumi}{(\alph{enumi})}
  \renewcommand{\labelitemi}{$\bullet$}
  \stdpackages\macros
}
\newcommand{\foga}{
  \documentclass{article} 
  \stdpackages\macros
  \usepackage{blabla/tex/foga-02}
  \usepackage{blabla/tex/chicago}
  \bibliographystyle{blabla/tex/foga-chicago}
}
\newcommand{\book}[2]{
  \documentclass[#1pt,twoside,fleqn]{book}
  \newenvironment{abstract}{\begin{rblock}{\bf Abstract.~}\small}{\end{rblock}}
  \stdstyle{#2}
  \macros
}

\newcommand{\foils}[1]{
  \documentclass[12pt,fleqn]{article}
  \stdstyle{#1}
  \voffset -1cm  \addtolength{\textheight}{2cm}
  \renewcommand{\footskip}{2cm}
  \macros
  
\begin{document}
  \large
}
\newcommand{\slides}{
  \documentclass{article}
  \stdpackages
  \renewcommand{\baselinestretch}{1}
  \renewcommand{\arraystretch}{1.8}

  \usepackage{geometry}
  \geometry{
    a4paper,landscape,dvips,
    headheight=15mm,
    headsep=20mm,
    footskip=0mm,
    hdivide={20mm,*,20mm},vdivide={5mm,*,5mm},
    twosideshift=0mm}

  \setlength{\columnsep}{30mm}
  \parindent 0ex
  \parskip 0ex
  \setlength{\itemsep}{8ex}

  \pagestyle{fancy}
  \renewcommand{\headrulewidth}{1pt}
  \renewcommand{\footrulewidth}{0pt}
  \renewcommand{\labelenumi}{\textbf{\arabic{enumi}.}~~}
  \newcommand{\theslide}{~}
  \newcommand{\theauthor}{Marc Toussaint}
  \rhead{}
  \lhead{{{\Huge\textsf{\quad\theslide}}\\}}
  \rfoot{\thepage}

  \newcommand{\inverted}{
    \definecolor{main}{rgb}{1,1,1}
    \color{main}
    \pagecolor{black}
  }

  \macros
}

\newcommand{\firstslide}[1]{\renewcommand{\theslide}{#1}}
\newcommand{\newslide}[1]{\onecolumn\renewcommand{\theslide}{#1}}
\newcommand{\newslidetwo}[1]{\twocolumn\renewcommand{\theslide}{#1}}


\author{Marc Toussaint}

\newcommand{\inilogo}{
}

\newcommand{\addressCologne}{
  Institute for Theoretical Physics\\
  University of Cologne\\
  50923 K\"oln---Germany\\
  {\tt mt@thp.uni-koeln.de}\\
  {\tt www.thp.uni-koeln.de/\~{}mt/}
}

\newcommand{\homepage}{\texttt{www.neuroinformatik.rub.de/PEOPLE/mt/}}
\newcommand{\email}{\textrm{mt@neuroinformatik.rub.de}}
\newcommand{\phone}{+49-234-32-27974}

\newcommand{\address}{
  Institut f\"ur Neuroinformatik,
  Ruhr-Universit\"at Bochum, ND 04,
  44780 Bochum---Germany
}
\newcommand{\Address}{
  Adaptive Systems Group -- Institut f\"ur Neuroinformatik\\
  Ruhr-Universit\"at Bochum, ND 04\\
  44780 Bochum---Germany
}

\newcommand{\published}{}

\newcommand{\mytitle}{
  \thispagestyle{empty}
  \hrule height2pt
  \begin{list}{}{\leftmargin2ex \rightmargin2ex \topsep2ex }\item[]
    {\huge\bf \thetitle}
  \end{list}
  \begin{list}{}{\leftmargin7ex \rightmargin7ex \topsep0ex }\item[]
    Marc Toussaint \quad\today

    {\small\it \address}
  \end{list}
  \vspace{2ex}
  \hrule height1pt
  \vspace{5ex}
  \renewcommand{\mytitle}{\chapter{\thetitle}}
}
\renewcommand{\mytitle}{
  \thispagestyle{empty}
  \mbox{~}
  \begin{list}{}{\leftmargin4ex \rightmargin4ex \topsep10ex }\item[]
    \begin{center}
      {\LARGE \thetitle \\}

      \vspace{8ex}
      {\large \theauthor}

      \vspace{1ex}
      {\footnotesize\sl \address}
      {\footnotesize \email}

      {\footnotesize\today}

      \vspace{1ex}
      {\small \published}
    \end{center}
  \end{list}
  \renewcommand{\mytitle}{\chapter{\thetitle}}
}


\newcommand{\subsec}[1]{
  \addtocontents{toc}{
      \protect\vspace*{-1.5ex}\protect\hspace*{26mm}
      \protect\begin{minipage}[t]{110mm}\protect\footnotesize\protect\textsf{#1}\protect\end{minipage}
      \protect\par
  }
  \begin{rblock}\it #1\end{rblock}\medskip\noindent
}
\newcommand{\tocsep}{
  \addtocontents{toc}{\protect\bigskip}
}
\newcommand{\Chapter}[1]{
\chapter*{#1}\thispagestyle{empty}
\addcontentsline{toc}{chapter}{\protect\numberline{}#1}
}
\newcommand{\Section}[1]{
  \section*{#1}
  \addcontentsline{toc}{section}{\protect\numberline{}#1}
}
\newcommand{\Subsection}[1]{
  \subsection*{#1}
  \addcontentsline{toc}{subsection}{\protect\numberline{}#1}
}

\newcommand{\content}[1]{
}

\newcommand{\sepline}{
  \begin{center} \begin{picture}(200,0)
    \line(1,0){200}
  \end{picture}\end{center}
}

\newcommand{\horline}{\hrule}

\newcommand{\sepstar}{
  \begin{center} {\vspace{0.5ex}*} \end{center}\vspace{-1.5ex}\noindent
}

\newcommand{\partsection}[1]{
  \vspace{5ex}
  \centerline{\sc\LARGE #1}
  \addtocontents{toc}{\contentsline{section}{{\sc #1}}{}}
}

\newcommand{\intro}[1]{\textbf{#1}\index{#1}}

\newcounter{parac}
\newcommand{\para}{\noindent\refstepcounter{parac}{\bf [{\roman{parac}}]}~~}
\newcommand{\Pref}[1]{[\emph{\ref{#1}}\,]}

\newenvironment{items}{
\begin{list}{}{\leftmargin1ex \topsep-\parskip}
\item[]
}{
\end{list}
}

\newenvironment{block}[1][]{{\noindent\bf #1}
\begin{list}{}{\leftmargin\blockindent \topsep-\parskip}
\item[]
}{
\end{list}
}

\newenvironment{rblock}{
\begin{list}{}{\leftmargin\blockindent \rightmargin\blockindent \topsep-\parskip}
\item[]
}{
\end{list}
}

\newenvironment{algorithm}{
\begin{list}{\raisebox{.3ex}{\footnotesize\bf(\arabic{enumi})}}
{\usecounter{enumi} \leftmargin7ex \rightmargin7ex \labelsep1ex
  \labelwidth5ex \topsep0ex \parsep.5ex \itemsep0pt}
}{
\end{list}
}

\newenvironment{keywords}{\paragraph{Keywords}\begin{rblock}\small}{\end{rblock}}

\newenvironment{colpage}{
\addtolength{\columnwidth}{-3ex}
\begin{minipage}{\columnwidth}
\vspace{.5ex}
}{
\vspace{.5ex}
\end{minipage}
}

\newenvironment{enum}{
\begin{list}{}{\leftmargin3ex \topsep0ex \itemsep0ex}
\item[\labelenumi]
}{
\end{list}
}

\newenvironment{cramp}{
\begin{quote} \begin{picture}(0,0)
        \put(-5,0){\line(1,0){20}}
        \put(-5,0){\line(0,-1){20}}
\end{picture}
}{
\begin{picture}(0,0)
        \put(-5,5){\line(1,0){20}}
        \put(-5,5){\line(0,1){20}}
\end{picture} \end{quote}
}


\newcommand{\macros}{
  \newcommand{\0}{{\hat 0}}
  \newcommand{\1}{{\hat 1}}
  \newcommand{\2}{{\hat 2}}
  \newcommand{\3}{{\hat 3}}
  \newcommand{\5}{{\hat 5}}

  \renewcommand{\a}{\ensuremath\alpha}
  \renewcommand{\b}{\beta}
  \renewcommand{\c}{\gamma}
  \renewcommand{\d}{\delta}
    \newcommand{\D}{\Delta}
    \newcommand{\e}{\epsilon}
    \newcommand{\g}{\gamma}
    \newcommand{\G}{\Gamma}
  \renewcommand{\l}{\lambda}
  \renewcommand{\L}{\Lambda}
    \newcommand{\m}{\mu}
    \newcommand{\n}{\nu}
    \newcommand{\N}{\nabla}
  \renewcommand{\k}{\kappa}
  \renewcommand{\o}{\omega}
  \renewcommand{\O}{\Omega}
    \newcommand{\p}{\phi}
    \newcommand{\ph}{\varphi}
  \renewcommand{\P}{\Phi}
  \renewcommand{\r}{\varrho}
    \newcommand{\s}{\sigma}
    \newcommand{\Si}{\Sigma}
  \renewcommand{\t}{\theta}
    \newcommand{\T}{\Theta}
  \renewcommand{\v}{\vartheta}
    \newcommand{\x}{\xi}
    \newcommand{\X}{\Xi}
    \newcommand{\Y}{\Upsilon}

  \renewcommand{\AA}{{\cal A}}
    \newcommand{\GG}{{\cal G}}
  \renewcommand{\SS}{{\cal S}}
    \newcommand{\TT}{{\cal T}}
    \newcommand{\EE}{{\cal E}}
    \newcommand{\FF}{{\cal F}}
    \newcommand{\HH}{{\cal H}}
    \newcommand{\II}{{\cal I}}
    \newcommand{\KK}{{\cal K}}
    \newcommand{\LL}{{\cal L}}
    \newcommand{\MM}{{\cal M}}
    \newcommand{\NN}{{\cal N}}
    \newcommand{\CC}{{\cal C}}
    \newcommand{\OO}{{\cal O}}
    \newcommand{\PP}{{\cal P}}
    \newcommand{\QQ}{{\cal Q}}
    \newcommand{\RR}{{\cal R}}
    \newcommand{\UU}{{\cal U}}
    \newcommand{\YY}{{\cal Y}}
    \newcommand{\SOSO}{{\cal SO}}
    \newcommand{\GLGL}{{\cal GL}}

  \newcommand{\NNN}{{\mathbb{N}}}
  \newcommand{\ZZZ}{{\mathbb{Z}}}
  \newcommand{\RRR}{{\mathbb{R}}}
  \newcommand{\CCC}{{\mathbb{C}}}
  \newcommand{\one}{{{\bf 1}}}
  \newcommand{\eee}{\text{e}}

  \renewcommand{\[}{\Big[}
  \renewcommand{\]}{\Big]}
  \renewcommand{\(}{\Big(}
  \renewcommand{\)}{\Big)}
  \renewcommand{\|}{\big|}
  \newcommand{\<}{{\ensuremath\langle}}
  \renewcommand{\>}{{\ensuremath\rangle}}

  \newcommand{\Prob}{{\rm Prob}}
  \newcommand{\Aut}{{\rm Aut}}
  \newcommand{\cor}{{\rm cor}}
  \newcommand{\corr}{{\rm corr}}
  \newcommand{\cov}{{\rm cov}}
  \newcommand{\sd}{{\rm sd}}
  \newcommand{\tr}{{\rm tr}}
  \newcommand{\Tr}{{\rm Tr}}
  \newcommand{\id}{{\rm id}}
  \newcommand{\Gl}{{\rm Gl}}
  \newcommand{\lag}{\mathcal{L}}
  \newcommand{\inn}{\rfloor}
  \newcommand{\lie}{\pounds}
  \newcommand{\longto}{\longrightarrow}
  \newcommand{\speer}{\parbox{0.4ex}{\raisebox{0.8ex}{$\nearrow$}}}
  \renewcommand{\dag}{ {}^\dagger }
  \newcommand{\h}{{}^\star}
  \newcommand{\w}{\wedge}
  \newcommand{\too}{\longrightarrow}
  \newcommand{\To}{\Rightarrow}
  \newcommand{\Too}{\;\Longrightarrow\;}
  \newcommand{\ow}{\stackrel{\circ}\wedge}
  \newcommand{\feed}{\nonumber \\}
  \newcommand{\comma}{\; , \quad}
  \newcommand{\period}{\; . \quad}
  \newcommand{\del}{\partial}
  \newcommand{\point}{$\bullet~~$}
  \newcommand{\doubletilde}{
  ~ \raisebox{0.3ex}{$\widetilde {}$} \raisebox{0.6ex}{$\widetilde {}$} \!\!
  }
  \newcommand{\topcirc}{\parbox{0ex}{~\raisebox{2.5ex}{${}^\circ$}}}
  \newcommand{\topdot} {\parbox{0ex}{~\raisebox{2.5ex}{$\cdot$}}}
  \newcommand{\topddot} {\parbox{0ex}{~\raisebox{1.3ex}{$\ddot{~}$}}}
  \newcommand{\sym}{\topcirc}

  \newcommand{\half}{\frac{1}{2}}
  \newcommand{\third}{\frac{1}{3}}
  \newcommand{\fourth}{\frac{1}{4}}

  \newcommand{\ubar}{\underline}

  \renewcommand{\_}{\underset}
  \renewcommand{\^}{\overset}
  \renewcommand{\*}{\text{\footnotesize\raisebox{-.4ex}{*}{}}}
}

\newcommand{\RND}{{\SS}}
\newcommand{\IF}{\text{if }}
\newcommand{\AND}{\textsc{and }}
\newcommand{\OR}{\textsc{or }}
\newcommand{\XOR}{\textsc{xor }}
\newcommand{\NOT}{\textsc{not }}
\newcommand{\argmax}[1]{{\rm arg}\!\max_{#1}}
\newcommand{\argmin}[1]{\text{arg}\!\min_{#1}}
\newcommand{\ee}[1]{\ensuremath{\cdot10^{#1}}}
\newcommand{\sub}[1]{\ensuremath{_{\text{#1}}}}
\newcommand{\up}[1]{\ensuremath{^{\text{#1}}}}
\newcommand{\kld}[2]{D\big(#1\,\big|\!\big|\,#2\big)}
\newcommand{\sprod}[2]{\big<#1\,,\,#2\big>}
\newcommand{\End}{\text{End}}
\newcommand{\txt}[1]{\quad\text{#1}\quad}


\newcommand{\anchor}[3]{\begin{picture}(0,0)\put(#1,#2){#3}\end{picture}}
\newcommand{\pagebox}{\begin{picture}(0,0)\put(-3,-34){
\framebox[\textwidth]{\rule[-\textheight]{0pt}{0pt}}}
\end{picture}}

\newcommand{\pathmt}{./}
\newcommand{\basepath}{./}
\newcommand{\setpath}[1]{\renewcommand{\pathmt}{#1}\renewcommand{\basepath}{#1}}
\newcommand{\pathinput}[2]{
  \renewcommand{\pathmt}{\basepath #1}
  \input{\pathmt #2} \renewcommand{\pathmt}{\basepath}}

\newcommand{\hide}[1]{{\tt[hide:~}{\footnotesize\sf
    #1}{\tt]}\message{^^JHIDE--Warning!^^J}}
\newcommand{\Hide}{\renewcommand{\hide}[1]{\message{^^JHIDE--Warning (hidden)!^^J}}}
\newcommand{\todo}[1]{{\tt[TODO: #1]}\message{^^JTODO--Warning: #1^^J}}
\newcommand{\Todo}{\renewcommand{\todo}[1]{\message{^^JTODO--Warning (hidden)!^^J}}}
\newcommand{\header}{\begin{document}\maketitle\cleardefs}
\newcommand{\contents}{{\tableofcontents}\renewcommand{\contents}{}}
\newcommand{\footer}{\small\bibliography{blabla/tex/bibs}\end{document}}


\article{10}{1.1}
\bibliographystyle{unsrt}
\Hide

\title{
Learning a world model and planning with a self-organizing, dynamic neural system
}

\hide{
\author{
Marc Toussaint \\
\Address \\
\texttt{mt@neuroinformatik.rub.de}\\
}
}

\newcommand{\mazehome}{../../code/projects/SWM/}
\newcommand{\bs}{{\bar s}}
\newcommand{\bS}{{\bar S}}
\newcommand{\td}{{\!\dagger}}
\newcommand{\tdd}{{\!\dagger\!\dagger}}

\newcommand{\cond}[3]{
\left\{\begin{array}{c@{~~~~}c} #2 & \text{if } #1 \\ #3 & \text{otherwise}\end{array}\right.
}
\newcommand{\iiff}[2]{
\begin{description}\item[if #1 then] #2\end{description}
}

\header

\begin{abstract}
  We present a connectionist architecture that can learn a model of
  the relations between perceptions and actions and use this model for
  behavior planning. State representations are learned with a growing
  self-organizing layer which is directly coupled to a perception and
  a motor layer. Knowledge about possible state transitions is encoded
  in the lateral connectivity. Motor signals modulate this lateral
  connectivity and a dynamic field on the layer organizes a planning
  process. All mechanisms are local and adaptation is based on Hebbian
  ideas. The model is continuous in the action, perception, and time
  domain.
\end{abstract}

\section{Introduction}

Planning of behavior requires some knowledge about the consequences of
actions in a given environment. A \emph{world model} captures such
knowledge. It seems that the brain is capable of planning in a way
that involves a simulation of actions and their perceptual
consequences (see, e.g., Hesslow's \cite{hesslow:02} arguments for a
\emph{simulation theory of cognitive brain function}).  However, the
level of abstraction, the representation, on which such simulation
occurs is hardly the level of physical coordinates. A tempting
hypothesis is that the representations the brain uses for reasoning
and planning are particularly designed (by adaptation or evolution)
for \emph{just} this purpose. \hide{They are categorical and neglect
  details that are irrelevant for the consequences of actions; they
  represent exactly those properties of situations or objects that are
  relevant to judge the consequences of actions.} To address such
ideas we first need a basic model for how a connectionist architecture
can encoded a world model and how self-organization of inherent
representations is possible.

In the field of machine learning, world models are a standard approach
to handle behavior organization problems (for a comparison of
model-based approaches to the classical, model-free Reinforcement
Learning see \cite{majors-richards:97}). Our approach for a
\emph{connectionist world model} (CWM) builds on the classical notions
of a world model and is functionally similar to existing Machine
Learning approaches with self-organizing state space models
\cite{kroese-eecen:94,zimmer:96}\hide{appl:00}. It is able to grow
neural representations for different world states and to learn the
consequences of actions in terms of state transitions. It differs
though from classical approaches in some crucial points:
\begin{list}{$\bullet$}{\leftmargin3ex \parsep0ex}
\item The model is continuous in the action, the perception, as well
  as the time domain.
\item All mechanisms are based on local interactions. The adaptation
  mechanisms are largely derived from the idea of Hebbian plasticity.
  E.g., the lateral connectivity, which encodes knowledge about
  possible state transition, is adapted by a variant of the temporal
  Hebb rule and allows local adaptation of the world model to local
  world changes.
\item The coupling to the motor system is fully integrated in the
  architecture via a mechanism incorporating modulating synapses
  (comparable to shunting mechanisms).
\item The two dynamic processes on the CWM, the ``tracking'' process
  estimating the current state and the planning process (similar to
  Dynamic Programming), will be realized by activation dynamics on the
  architecture, incorporating in particular lateral interactions,
  inspired by neural fields \cite{amari:77}.
\end{list}

\hide{
of self

---we will introduce the CWM as a
connectionist way to represent and learn a partially observable Markov
decision process and to organize a ``state tracking'' and a ``behavior
planning'' dynamics on this representation (in analogy to tracking in
hidden Markov models and in analogy to dynamic programming,
respectively). On the other hand, it adopts many techniques from
typical connectionist approaches, in particular inspired by dynamic
processes neural fields and the various self-adaptation methods.
Specifically, our model will
\begin{itemize}
\item self-organize a state space representation, similar to how
  FuzzyARTMAPs \cite{carpenter-et-al:92} grow representations, as it
  has already been proposed by
  \cite{kroese-eecen:94,zimmer:96,appl:00} in the realm of machine
  learning;
\item implement a neural dynamics on this representation layer, driven
  by forward perceptual excitations as well as lateral, predictive
  excitations;
\item learn how actions influence lateral excitations (i.e., state
  transition probabilities) via a novel way of coupling a motor layer
  to the representation layer;
\item learn the weights of lateral connections (which correspond to
  probabilities of state transitions) by a variant of the temporal
  Hebb rule---what will allow quick adaptation of the model to world
  changes;
\item and realize behavior planning by another dynamic process on the
  representation layer, which is derived as a variant of Dynamic
  programming.
\end{itemize}
}

The outline of the paper is as follows: In the next section we
describe our architecture, the dynamics of activation and the
couplings to perception and motor layers. In section \ref{secV} we
introduce a dynamic process that generates, as an attractor, a
\emph{value field} over the layer which is comparable to a state value
function estimating the expected future return and allows for
goal-oriented behavior organization. The self-organization process and
adaptation mechanisms are described in section \ref{secAda}. We
demonstrate the features of the model on a maze problem in section
\ref{secDemo} and finally discuss the results and the model in general
terms.

\section{The model}\label{secModel}

\begin{figure}[t]\center
\input{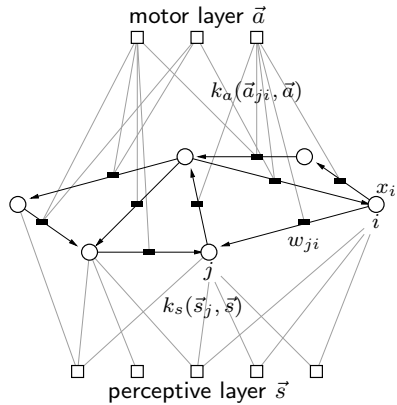}
\caption{\label{figCWM}
  Schema of the CWM architecture.}
\end{figure}

The core of the \emph{connectionist world model} (CWM) is a neural
layer which is coupled to a perceptual layer and a motor layer, see
figure \ref{figCWM}. Let us enumerate the units of the central layer
by $i=1,..,N$. Lateral connections within the layer may exist and we
denote a connection from the $i$-th to $j$-th unit by $(ji)$. E.g.,
``$\sum_{(ji)}$'' means ``summing over all existing connections
$(ji)$". To every unit we associate an activation $x_j \in \RRR$ which
is governed by the dynamics
\begin{align}\label{act}
\tau_x~ \dot x_j
=  - x_j + k_s(\vec s_j,\vec s)
   + \eta~ \sum_{(ji)} k_a(\vec a_{ji},\vec a)~ w_{ji}~ x_i ~,
\end{align}
which we will explain in detail in the following. First of all, $x_i$
are the time-dependent activations and the dot-notation $\tau_x~ \dot
x = F(x)$ means a time derivative which we algorithmically implemented
by a Euler integration step $x(t) = x(t-1) + \frac{1}{\tau_x}~
F(x(t-1))$.

The first term in (\ref{act}) induces an exponential relaxation while
the second and third terms are the inputs. $k_s(\vec s_j,\vec s)$ is
the forward excitation that unit $j$ receives from the perceptive
layer. Here, $\vec s_j$ is the codebook vector (receptive field) of
unit $j$ onto the perception layer which is compared to the current
stimulus $\vec s$ via the kernel function $k_s$. We will choose
Gaussian kernels as it is the case, e.g., for typical Radial Basis
function networks.

The third term, $\sum_{(ji)} k_a(\vec a_{ji},\vec a)~ w_{ji}~ x_i$,
describes the lateral interaction on the central layer.  Namely, unit
$j$ receives lateral input from unit $i$ iff there exists a connection
$(ji)$ from $i$ to $j$. This lateral input is weighted by the
connection's synaptic strength $w_{ji}$. Additionally there is another
term entering multiplicatively into this lateral interaction:
\emph{Lateral inputs are modulated depending on the current motor
  activation.} We chose a modulation of the following kind: To every
existing connection $(ji)$ we associate a codebook vector $\vec
a_{ji}$ onto the motor layer which is compared to the current motor
activation $\vec a$ via a Gaussian kernel function $k_a$. Due to the
multiplicative coupling, a connection contributes to lateral inputs
only when the current motor activation ``matches'' the codebook vector
of this connection. The modulation of information transmission by
multiplicative or divisive interactions is a fundamental principle in
biological neural systems \cite{phillips-singer:97}. One example is
shunting inhibition where inhibitory synapses attach to regions of the
dentritic tree near to the soma and thereby modulate the transmission
of the dentritic input \cite{abbott:91}. In our architecture, a
shunting synapse, receiving input from the motor layer, might attach
to only one branch of a (lateral) dentritic tree and thereby
multiplicatively modulate the lateral inputs summed up at this
subtree.

\hide{
\begin{figure}[t]\center
\input{figs/bio}
\caption{\label{figShunt}
  A shunting synapse (black bar) connecting to a dentritic branch. The
  square unit corresponds to our motor units. In effect, the coupling
  is multiplicative: All the excitation summed up at the dentritic
  branch is transmitted or not transmitted to the post-synaptic neuron
  depending on the shunting synapse.}
\end{figure}

Let us briefly consider the biological plausibility of such a
multiplicative coupling. In nature, synapses do not have a receptive
field to a large number of other units which can activate or
deactivate synaptic transfer. However, there exist so-called shunting
synapses, i.e., inhibitory synapses that attach to regions of the
dentritic tree near to the soma and there induce a shunting inhibition
which couples in a multiplicative (or rather divisive) way to the
incoming excitation current of that dentritic tree (see in particular
\cite{abbott:91}, equation (3.17); but also
\cite{torre-poggio:78,ermentrout:98,chance-abbott-reyes:02}). Assume
that such a shunting synapse is connected to only one branch of a
dentritic tree, see figure \ref{figShunt}. This shunting synapse may
induce a multiplicative shunting inhibition of the whole branch such
that all excitations from ``normal'' synapses entering this branch are
inhibited and cannot anymore excite the post-synaptic neuron. In this
way, the shunting synapse effectively switches on or off a whole set
of normal excitatory or inhibitory synapses while not affecting the
rest of synapses connected to other dentritic branches. Referring to
our architecture, a shunting synapse could be connected to a motor
unit such that the current motor activation determines which synapses
linking to the dentritic tree are effectively switched on or off.
}

For the following it is helpful if we briefly discuss a certain
relation between equation (\ref{act}) and a classical probabilistic
approach. Let us assume normalized kernel functions
\begin{align*}
k_s(\vec s_j,\vec s)
= \frac{1}{\sqrt{2~\pi}~ \s_s}\exp\frac{-(\vec s_j-\vec s)^2}{2~ \s_s^2} \comma
k_a(\vec a_{ji},\vec a)
= \frac{1}{\sqrt{2~\pi}~ \s_a}\exp\frac{-(\vec a_{ji}-\vec a)^2}{2~ \s_a^2} ~.
\end{align*}
These kernel functions can directly be interpreted as probabilities:
$k_s(\vec s_j,\vec s)$ represents the probability $P(\vec s|j)$ that
the stimulus is $\vec s$ if $j$ is active, and $k_a(\vec a_{ji},\vec
a)$ the probability $P(\vec a|j,i)$ that the action is $\vec a$ if
a transition $i \to j$ occurred. As for typical hidden Markov models
we may derive the \emph{prior} probability distribution $P(j|\vec
a)$, given the action:
\begin{align*}
P(j|\vec a,i)
&= \frac{P(\vec a|j,i)~ P(j|i)}{P(\vec a|i)}
 = k_a(\vec a_{ji},\vec a)~ \frac{P(j|i)}{P(\vec a|i)} \comma\\
P(j|\vec a)
&= \sum_{i} k_a(\vec a_{ji},\vec a)~ \frac{P(j|i)}{P(\vec a|i)}~ P(i) ~.
\end{align*}
$P(\vec a|i)$ can be computed by normalizing $P(\vec a|j,i)~ P(j|i)$
over $j$ such that $\sum_j P(j|\vec a,i) = 1$. What we would like to
mention here is that in equation (\ref{act}), the lateral input\\
$\sum_{(ji)} k_a(\vec a_{ji},\vec a)~ w_{ji}~ x_i$ can be compared to
the prior $P(j|\vec a)$ under the assumption that $x_i$ is
proportional to $P(i)$ and if we have an adaptation mechanism for
$w_{ji}$ which converges to a value proportional to $P(j|i)$ and which
also ensures normalization, i.e., $\sum_j k_a(\vec a_{ji},\vec a)~
w_{ji} = 1$ for all $i$ and $\vec a$. This insight will help to judge
some details of the next two section. The probabilistic interpretation
can be further exploited, e.g., comparing the input of a unit $j$ (or,
in the quasi-stationary case, $x_j$ itself) to the \emph{posterior}
and deriving theoretically grounded adaptation mechanisms. But this is
not within the scope of this paper.

\hide{

\section{Comparing the CWM to a Bayesian model}\label{secEM}
\renewcommand{\td}{{}}

A probabilistic interpretation of the CWM is possible if we assume
normalized kernel functions, e.g.,
\begin{align*}
k_s(\vec s_j,\vec s)
= \frac{1}{\sqrt{2~\pi}~ \s_s}\exp\frac{-(\vec s_j-\vec s)^2}{2~ \s_s^2} \comma
k_a(\vec a_{ji},\vec a)
= \frac{1}{\sqrt{2~\pi}~ \s_a}\exp\frac{-(\vec a_{ji}-\vec a)^2}{2~ \s_a^2} ~.
\end{align*}
These kernel functions directly represent probabilities, more
precisely, $k_s(\vec s_j,\vec s)$ represents the probability $P(\vec
s|j)$ that the stimulus is $\vec s$ if $j$ is active, and $k_a(\vec
a_{ji},\vec a)$ the probability $P(\vec a|j,i^\td)$ that the action is
$\vec a$ if a transition $i \to j$ occurred. As for typical hidden
Markov models we may derive the (\emph{prior}) probability
distribution $P(j|\vec a)$ given the action, and the
(\emph{posterior}) probability distribution $P(j|\vec s,\vec a)$ given
in addition the information on the received stimulus:
\begin{align*}
P(j|\vec a,i^\td)
&= \frac{P(\vec a|j,i^\td)~ P(j|i^\td)}{P(\vec a|i^\td)}
 = \frac{k_a(\vec a_{ji},\vec a)~ P(j|i^\td)}{P(\vec a|i^\td)} ~.
\end{align*}
$P(\vec a|i^\td)$ can be computed by normalizing $P(\vec a|j,i^\td)~
P(j|i^\td)$ over $j$ such that $\sum_j P(j|\vec a,i^\td) = 1$. Given
$P(i^\td)$ we get
\begin{align*}
P(j|\vec a)
&= \sum_{i} k_a(\vec a_{ji},\vec a)~ \frac{P(j|i^\td)}{P(\vec a|i^\td)}~ P(i^\td) ~.
\end{align*}
Let us compare this prior to our definition of the lateral input $\bar
x_j = \sum_{(ji)} k_a(\vec a_{ji},\vec a)~ w_{ji}~ x_i$. Under the
assumption that $x_i$ is proportional to $P(i^\td)$ and if we have an
adaptation mechanism for $w_{ji}$ which converges to a value
proportional to $P(j|i^\td)$ and which also ensures normalization,
i.e., $\sum_j k_a(\vec a_{ji},\vec a)~ w_{ji} = 1$ for all $i$ and
$\vec a$, then $\bar x_j$ is indeed proportional to the prior
$P(j|\vec a)$.
\hide{
\begin{align*}
&\sim& \bar x_j = \sum_{(ji)} k_a(\vec a_{ji},\vec a)~ w_{ji}~ x_i^\td\\
&\sim& \dot x_j = \frac{1}{\tau} \[ k_s(\vec x_j,\vec s) + \eta~ \bar x_j - x_j \]
\end{align*}

Now, if we define that in the probabilistic model we always chose a
connection's weight equal to
\begin{align}\label{wPro}
w_{ji} = \frac{P(j|i^\td)}{\sum_k x_{ki}~ P(k|i^\td)}
\end{align}
and, if we take the prior belief probability $P(i^\td)$ equal to the
posterior of the previous time step $P(i^\td|\vec s^\td,\vec a^\td)$,
we retrieve equation (\ref{latEx}),
\begin{align*}
\bar x_j
&= P(j|\vec a)
 = \sum_{i} x_{ji}~ w_{ji}~ x_i^\td ~.
\end{align*}
}
Concerning the posterior, it holds
\begin{align*}
P(j|\vec s,\vec a)
&= \frac{P(\vec s|j,\vec a)~ P(j|\vec a)}{P(\vec s|\vec a)}
&\equiv \frac{P(\vec s|j)~ P(j|\vec a)}{P(\vec s|\vec a)} ~.
 = \frac{1}{P(\vec s|\vec a)}~ k_s(\vec x_j,\vec s)~ P(j|\vec a)
\end{align*}
Again, $P(\vec s|\vec a)$ can be calculated as the normalization of
$P(\vec s|j)~ P(j|\vec a)$ over $j$ such that $\sum_j P(j|\vec s,\vec
a) = 1$. Comparing this posterior to the \emph{fixed point} $k_s(\vec
s_j,\vec s) + \eta~ \bar x_j$ of our activation dynamics (\ref{act})
exhibits the difference between our model and a straight-forward
hidden Markov model: First, in our approach the forward and lateral
excitations are coupled additively instead of multiplicatively. And
second, time continuity of our model may become important if the model
does not operate quasi-stationary.

}

\section{The dynamics of planning}\label{secV}

To organize goal-oriented behavior we assume that, in parallel to the
activation dynamics (\ref{act}), there exists a second dynamic process
which can be motivated from classical approaches to Reinforcement
Learning \cite{bertsekas-tsitsiklis:96,sutton-barto:98}. Recall the
\emph{Bellman equation}
\begin{align}\label{fixV}
V^*_\pi (i)
 = \sum_a \pi(a|i) \sum_j P(j|i,a) \[ r(j) + \g~  V^*_\pi (j) \] ~,
\end{align}
yielded by the \emph{expectation} $V^*(i)$ of the discounted
\emph{future} return $R(t)=\sum_{\tau=1}^\infty \g^{\tau\!-\!1}~
\r(t\!+\!\tau)$, which yields $R(t)=\r(t\!+\!1) + \g~ R(t\!+\!1)$,
when situated in state $i$. Here, $\g$ is the discount factor and we
presumed that the received rewards $\r(t)$ actually depend only on the
state and thus enter equation (\ref{fixV}) only in terms of the reward
function $r(i)$ (we neglect here that rewards may directly depend on
the action). Behavior is described by a stochastic policy $\pi(a|i)$,
the probability of executing action $a$ in state $i$. Knowing the
property (\ref{fixV}) of $V^*$ it is straight-forward to define a
recursion algorithm for an approximation $V$ of $V^*$ such that $V$
converges to $V^*$. This recursion algorithm is called \emph{Value
  Iteration} and reads
\begin{align}\label{defDP}
\tau_v~ \D V_\pi (i)
& = - V_\pi (i)
  + \sum_a \pi(a|i) \sum_j P(j|i,a) \big[ r(j) + \g~ V_\pi (j) \big] ~,
\end{align}
with a ``reciprocal learning rate'' or time constant $\tau_v$. Note
that (\ref{fixV}) is the fixed point equation of (\ref{defDP}).

\hide{

The computational complexity of Value Iteration is a problem: The
iteration has to be done for all function components $V(i)$ (i.e., for
all $i$), summations goes over all $a$ and all $j$. Thus, methods
to optimize this iteration procedure have been developed and are
commonly applied (\emph{Real-Time Dynamic Programming}
\cite{barto-bradtke-singh:95}; \emph{prioritized sweeping}
\cite{moore-atkeson:1993}).

In classical, model-free Reinforcement Learning, this algorithm is
realized only approximately: Not all the summations over probabilities
are taken into account since, in model-free approaches, these
probabilities are neither known nor approximated. Instead, only the
one single experience that was made (i.e., for on-policy or SARSA
learning, the actually occurred sample of the probabilities $\pi(a|i)$
and $P(j|i,a)$) is used for the update such that the summations drop
out:
\begin{align*}
\D V(i) = \frac{1}{\tau_v} \[ r(i) + \g~ V(j) - V(i) \]
\end{align*}
where $i$ and $j$ are the actual system states at time $t-1$ and $t$,
respectively.

}

The practical meaning of the state-value function $V$ is that it
quantifies how desirable and promising it is to reach a state $i$,
also accounting for future rewards to be expected. In particular, if
one knows the current state $i$ it is a simple and efficient rule of
behavior to choose that action $a$ that will lead to the neighbor
state $j$ with maximal $V(j)$ (the greedy policy). In that sense,
$V(i)$ provides a smooth gradient towards desirable goals. Note though
that direct Value Iteration presumes that the state and action spaces
are known and finite, and that the current state and the world model
$P(j|i,a)$ is known.

How can we transfer these classical ideas to our model? We suppose
that the CWM is given a goal stimulus $\vec g$ from outside, i.e., it
is given the command to reach a world state that corresponds to the
stimulus $\vec g$.  This stimulus induces a \emph{reward excitation}
$r_i = k_s(\vec s_i,\vec g)$ for each unit $i$. Now, besides the
activations $x_i$, we introduce another field over the CWM, the
\emph{value field} $v_i$, which is in analogy to the state-value
function $V(i)$. The dynamics is
\begin{align}\label{valuefield}
\tau_v~ \dot v_i =
& - v_i + r_i + \g~ \max_{(ji)}(w_{ji}~ v_j) ~,
\end{align}
and well comparable to (\ref{defDP}): A slight difference is that
$v_i$ estimates the ``current-plus-future'' reward $\r(t) + \g R(t)$
rather than the future reward only---in the upper notation this
corresponds to the slightly modified value iteration $\tau_v~ \D
V_\pi(i) = - V_\pi(i) + r(i) + \sum_a \pi(a|i) \sum_j P(j|i,a) \big[
\g~ V_\pi(j) \big]$. As it is commonly done for Value Iteration, we
assumed $\pi$ to be the greedy policy. More precisely, we considered
only that action (i.e., that connection $(ji)$) that leads to the
neighbor state $j$ with maximal value $w_{ji}~ v_j$. In effect, the
summations over $a$ as well as over $j$ can be replaced by a
maximization over $(ji)$. Finally we replaced the probability factor
$P(j|i,a)$ by $w_{ji}$---we will see in the next section how
$w_{ji}$ is learned and what it will converge to.

In practice, the value field will relax quickly to its fixed point
$v_i^* = r_i + \g~ \max_{(ji)}(w_{ji}~ v_j^*)$ and stay there if the
goal does not change and if the world model is not re-adapted (see the
experiments). The quasi-stationary value field $v_i$ together with the
current (typically non-stationary) activations $x_i$ allow the system
to generate a motor signal that guides towards the goal. More
precisely, the value field $v_i$ determines for every unit $i$ the
``best'' neighbor unit $k_i=\argmax{j}~ w_{ji}~ v_j$. The output motor
signal is then the activation average
\begin{align}\label{actSel}
\vec a = \sum_i x_i~ \vec a_{k_ii}
\end{align}
of the motor codebook vectors $\vec a_{k_ii}$ that have been learned
for the corresponding connections. Hence, the information flow between
the central layer and the motor system is in both ways: In the
``tracking'' process as given by equation (\ref{act}) the information
flows from the motor layer to the central layer: Motor signals
activate the corresponding connections and cause lateral, predictive
excitations. In the action selection process as given by equation
(\ref{actSel}) the signals flow from the central layer back to the
motor layer to induce the motor activations that should turn
predictions into reality.

Depending on the specific problem and the motor system, a
post-processing of the motor signal $\vec a$, e.g.\ a competition
between contradictory motor units, might be necessary. In our
experiments we will have two motor units and will always normalize the
2D vector $\vec a$ to unit length.

\renewcommand{\td}{{\!\dagger}}

\section{Self-organization and adaptation}\label{secAda}

The self-organization process of the central layer combines techniques
from standard self-organizing maps \cite{malsburg:73,kohonen:95} and
their extensions w.r.t.\ growing representations
\cite{carpenter-et-al:92,fritzke:95}\hide{,bednar-kelkar-miikkulainen:02}
and the learning of temporal dependencies in lateral connections
\cite{bishop-hinton-strachan:97,somervuo:99,wiemer:03}\hide{euliano-principe:99,varsta:02,klemm-alstroem:02,}.
The free variables of a CWM subject to adaptation are (1) the number
of neurons and the lateral connectivity itself, (2) the codebook
vectors $\vec s_i$ and $\vec a_{ji}$ to the perceptive and motor
layers, respectively, and (3) the weights $w_{ji}$ of the lateral
connections.  The adaptation mechanisms we propose are based on three
general principles: (1) the addition of units for representation of
novel states (\emph{novelty}), (2) the fine tuning of the codebook
vectors of units and connections (\emph{plasticity}), and (3) the
adaptation of lateral connections in favor of better prediction
performance (\emph{prediction}).

\paragraph{Novelty.}
Mechanisms similar to those of FuzzyARTMAPs \cite{carpenter-et-al:92}
or Growing Neural Gas \cite{fritzke:95} account for the insertion of
new units when novelty is detected. We detect novelty in a
straight-forward manner, namely when the difference between the actual
perception and the best matching unit becomes too large. To make this
detection more robust, we use a low-pass filter (leaky integrator). At
a given time, let $z$ be the best matching unit, $z = \argmax{i} x_i$.
For this unit we integrate the error measure $e_z$
\begin{align*}
\tau_e~ \dot e_z = - e_z + (1-k_s(\vec s_z,\vec s)) ~.
\end{align*}
We normalize $k_s(\vec s_z,\vec s)$ such that it equals $1$ in the
perfect matching case when $\vec s_z = \vec s$. Whenever this error
measure exceeds a threshold called \emph{vigilance}, $e_z>\n$, $\n \in
[0,1]$, we generate a new unit $j$ with the codebook vector equal to the
current perception, $\vec s_j = \vec s$, and a connection from the
last best matching unit $z^\td$ with the codebook vector equal to the
current motor signal, $\vec a_{jz^\td} = \vec a$. The errors of both,
the new and the old unit, are reset to zero, $e_z \gets 0$, $e_j = 0$.

\paragraph{Plasticity.}
We use simple Hebbian plasticity to fine tune the representations of
existing units and connections. Over time, the receptive fields of units and
connections become more and more similar to the average stimuli that
activated them. We use the update rules
\begin{align*}
\tau_s~ \topdot \vec s_z  =  - \vec s_z + \vec s \comma
\tau_a~ \topdot \vec a_{zz^\td}  =  - \vec a_{zz^\td} + \vec a ~,
\end{align*}
with learning time constants $\tau_s$ and $\tau_a$.

\paragraph{Prediction and a temporal Hebb rule.}
Although perfect prediction is not the actual objective of the CWM,
the predictive power is a measure of the correctness of the learned
world model and good predictive power is one-to-one with good behavior
planning. The first and simple mechanism to adapt the predictive power
is to grow a new lateral connection between two successive best
matching units $z^\td$ and $z$ if it does not yet exist. The new
connection is initialized with $w_{zz^\td}=1$ and $\vec a_{zz^\td}
=\vec a$. The second, more interesting mechanism addresses the
adaptation of $w_{ji}$ based on new experiences and can be motivated
as follows: The temporal Hebb rule strengthens a synapse if the pre-
and post-synaptic neurons spike in sequence, depending on the
inter-spike-interval, and is supposed to roughly describe LTP and LTD
(see, e.g.,\cite{dayan-abbott:01}). In a population code model, this
corresponds to a measure of correlation between the pre-synaptic and
the delayed post-synaptic activity. In our case we additionally have
to account for the action-dependence of a lateral connection. We do so
by considering the term $k_a(\vec a_{ji},\vec a)~ x_i$ instead of only
the pre-synaptic activity. As a measure of temporal correlation we
choose to relate this term to the \emph{derivative} $\dot x_j$ of the
post-synaptic unit instead of its delayed activation---this saves us
from specifying an ad-hoc ``typical'' delay and directly reflects
that, in equation (\ref{act}), lateral inputs relate to the derivative
of $x_j$. Hence, we consider the product $\dot x_j~ k_a(\vec
a_{ji},\vec a)~ x_i$ as the measure of correlation. Our concrete
implementation is a robust version of this idea:
\begin{align*}
\tau_w~ \dot w_{ji}
&= \k_{ji}~ [ c_{ji} - w_{ji}~ \k_{ji} ] \comma \text{where}\\
&\tau_\k~ \dot c_{ji}
= - c_{ji} + \dot x_j ~ k_a(\vec a_{ji},\vec a)~ x_i \comma
\tau_\k~ \dot\k_{ji}
= - \k_{ji} + k_a(\vec a_{ji},\vec a)~ x_i  ~.
\end{align*}
Here, $c_{ji}$ and $\k_{ji}$ are simply low-pass filters of $\dot x_j~
k_a(\vec a_{ji},\vec a)~ x_i$ and of $k_a(\vec a_{ji},\vec a)~ x_i$.
The term $w_{ji}~ \k_{ji}$ ensures convergence (assuming quasi static
$c_{ji}$ and $\k_{ji}$) of $w_{ji}$ towards $c_{ji} \big/ \k_{ji}$.
The time scale of adaptation is modulated by the recent activity
$\k_{ji}$ of the connection.

\hide{
from our comparison of $w_{ji}$ with
$P(j|i)$: Instead of the correct probabilistic term $P(j,i)/P(i)$ we
consider $ \<\dot x_j(t)~ x_i(t)\> \big/ \<x_i\>$ noting that in
equation (\ref{act}), lateral excitation constitute the \emph{increase}
of $x_j$ and not $x_j$ absolutely. Here, $\< \cdot \>$ corresponds to
a time averaging that we realize by simple relaxation schemes:
\begin{align*}
\<\dot
x_j(t)~ x_i(t)\> \sim \dot c_{ji} = \frac{1}{\tau_\k} \[ \dot x_j ~
k_a(\vec a_{ji},\vec a)~ x_i - c_{ji} \] \comma \<x_i\> \sim \dot
\k_{ji} = \frac{1}{\tau_\k} \[ k_a(\vec a_{ji},\vec a)~ x_i - \k_{ji}
\] ~.
\end{align*}
Our adaptation dynamics
\begin{align*}
\dot w_{ji}
= \frac{\k_{ji}}{\tau_w}~ \[ c_{ji} - w_{ji}~ \k_{ji} \] \comma
\end{align*}
of $w_{ji}$ then converges (assuming quasi static $c_{ji}$ and
$\k_{ji}$) towards $c_{ji} \big/ \k_{ji}$ while the time scale of
adaptation is modulated by the recent activity $\k_{ji}$ of the
connection.

as a variant of the temporal Hebb rule. In
\cite{dayan-abbott:01} it is given as
\begin{align}\label{Hada}
\dot w_{ji}
= \frac{1}{\tau_w}~ \int_0^\infty\!\!\! d\tau
  ~ \[ H(\tau)~ x_j^{(t)}~ x_i^{(t-\tau)} + H(-\tau)~
  x_j^{(t-\tau)}~ x_i^{(t)} \] ~.
\end{align}
The term in the first brackets measures the temporal correlation
between the activation of the pre- and post-synaptic units. The
function $H(\tau)$ determines the rate of synaptic plasticity
depending on the ``inter-activity'' interval $\tau$ and is supposed to
roughly describe the LTP and LTD at the synapse. If the pre-activated
neuron is typically activated shortly before the post-synaptic then
the synapse is strengthened; in the opposite case it is weakened.

}

\hide{
Based on
the probabilistic interpretation given in section \ref{secEM} It is
possible to derive rigorous adaptation mechanisms of the EM kind, but
this is work in progress, see appendix \ref{secEM}. At this place let
me propose some heuristics that are based on the detection of a wrong
prediction (which can also be implemented by a temporal Hebb rule, see
below). Namely, if a wrong prediction is detected, weaken the edge
that lead to this wrong prediction \emph{and} the pre-synaptic unit of
this edge. The idea is that the state classification done by the
pre-synaptic unit was obviously no good for predictability; probably
another belief state with similar or same receptive field would have
been more appropriate. If, after several adaptations, an edge becomes
too weak it is deleted; and if a unit becomes too week, it is
duplicated.  We call the latter \emph{belief state duplication}. In
principle, this is a way to handle stimulus ambiguity in \emph{partial
  observable} MDPs. After some time, the two duplicated neurons might
still have the same or similar receptive field, but their
\emph{lateral} connectivity should be different such that their
pre-activation depends on the temporal context of that world state and
allows to distinguish between them. The more pre-activated unit will
win the competition if the ambiguous stimulus occurs.

In detail, the mechanisms accounting for this adaptation are the
following. Again, to simplify the adaptation mechanisms, we focus on
the best matching units $z$ and $z^\td$ in the current and previous
time step, respectively.
\begin{itemize}
\item First, if a connection from $z^\td$ to $z$ does not yet exist,
  then create it. Initialize it with $w_{zz^\td} = 1$ and
  $\vec a_{zz^\td} =\vec a$.

\item We define a winner-takes-all prediction unit $(p)$ as follows:
\begin{align*}
&p = \cond{\exists_i:~ \bar x_i > .5}{\argmax{i} \bar x_i}{z^\td}
  \end{align*}
  We use this only unit to detect mispredictions, i.e., when $z \not= p$.
  
\item Depending on the correctness of prediction we adapt the weight
  of the connection $(pz^\td)$ that was (among others) responsible for
  the prediction. In the discrete environment demonstration we set the
  weight immediately to zero in case of a misprediction. In the
  continuous case we use a smoother adaptation mechanism which is
  inspired from the temporal Hebb rule; for details see appendix
  \ref{secAlgo}. If the weight of a connection goes below a threshold,
  we delete this connection.

\item In a similar way, we accumulate the prediction errors that were
  caused by the activation of a pre-synaptic neuron via a leaky
  integrator. The pre-synaptic neuron was $z^\td$ and we integrate
  mispredictions as \todo{i oder z}
\begin{align*}
\dot l_i = \frac{1}{\tau_l} \[\sum_{(ji)} [j\not=z]~
    k_a(\vec a_{ji},\vec a)~ x_i^\td - l_i \]
  \end{align*}
  Here, the bracket [\textit{expression}] is 1 if the expression is
  true and 0 otherwise. If this error measure exceeds a threshold,
  $l_{z^\td} > \m$, $\m \in [0,1]$, we duplicate the neuron (belief
  state duplication). The new neuron $j$ inherits the receptive field,
  $\vec s_j = \vec s_{z^\td}$ and all the lateral input and output
  connections (which are also duplicated). $e_{z^\td} \gets 0$, $e_j =
  0$.
\end{itemize}
}

\section{Experiments}\label{secDemo}

\begin{figure}[t]\center
\parbox{1mm}{\large a}  \raisebox{-40mm}{\input{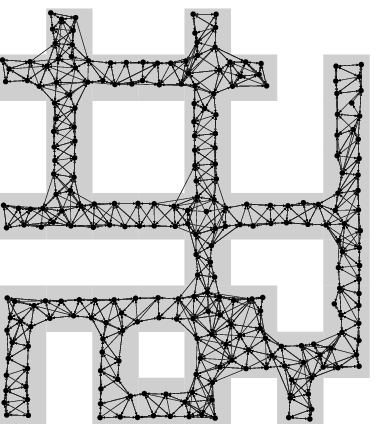}}~
\parbox{1mm}{\large b} \raisebox{-40mm}{\input{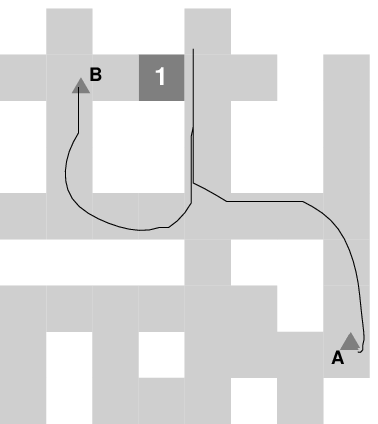}}~
\parbox{1mm}{\large c}\raisebox{-40mm}{\input{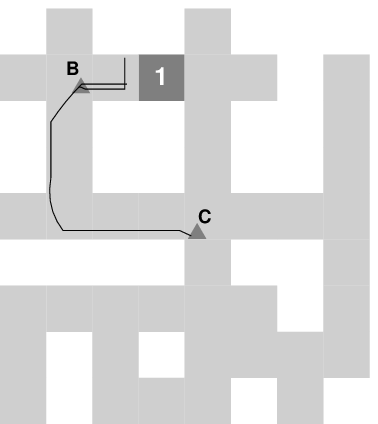}}\\[1ex]
\parbox{1mm}{\large d} \raisebox{-40mm}{\input{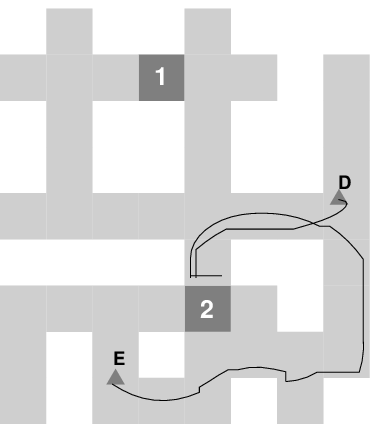}}~
\parbox{1mm}{\large e}  \raisebox{-40mm}{\input{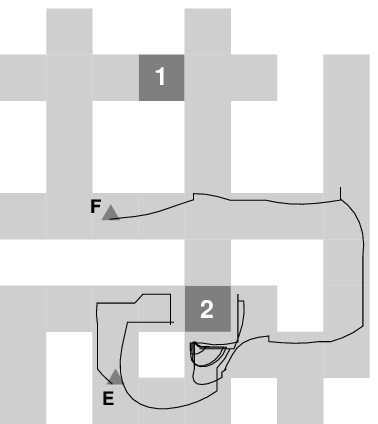}}~
\parbox{1mm}{\large f} \raisebox{-40mm}{\input{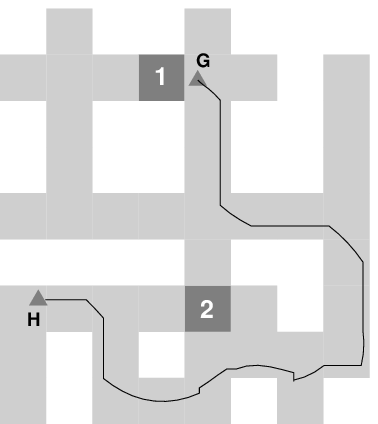}}
\caption{\label{figMaze}
  The CWM on a maze problem: (a) the outcome of self-organization;
  (b-c) agent movements from goal {\sf A} to {\sf B} to {\sf C}, here, the trespass 1
  was blocked and requires readaptation of the world model; (d-f)
  agent movements that demonstrate adaptation to a second blockade.
  Please see the text for more explanations. }
\end{figure}

To demonstrate the functionality of the CWM we consider a simple maze
problem. The parameters we used are\\[2mm]
\centerline{\begin{tabular}{ccccccccccc} $\tau_x$ & $\eta$ & $2\,
    \s_s^2$ & $2\, \s_a^2$ & $\tau_v$ & $\g$ & $\tau_e$ &
    $\tau_s$ & $\tau_a$ & $\tau_w$ & $\tau_\k$ \\[1mm]
    \hline\\[-3mm]
    2 & 0.1 & 0.01 & 0.5 & 2 & 0.8 & 10 & 20 & 5 & 10 & 100
\end{tabular}~.}\\[2mm]
Figure \ref{figMaze}a displays the geometry of the maze. The ``agent''
is allowed to move continuously in this maze. The motor signal is
2-dimensional and encodes the forces $\vec f$ in x- and y-directions;
the agent has a momentum and friction according to $\topddot\vec x =
0.2\, (\vec f- \topdot\vec x)$. As a stimulus, the CWM is given the 2D
position $\vec x$.

Figure \ref{figMaze}a also displays the (lateral) topology of the
central layer after 30\,000 time steps of self-organization, after
which the system becomes quasi-stationary.  The model is learned from
scratch, initialized with one random unit.  During this first phase,
behavior planning is switched off and the maze is explored with a
random walk that changes its direction only with probability 0.1 at a
time. In the illustration, the positions of the units correspond to
the codebook vectors that have been learned.  The directedness and the
codebook vectors of the connections can not displayed.
\hide{ Analysis of this CWM reveals that the codebook vectors learned
  for each connection approximate the geometric direction of the
  connection: The average deviation is 25 degrees.  }

After the self-organization phase we switched on behavior planning. A
goal stimulus corresponding to a random position in the maze is given
and changed every time the agent reaches the goal. Generally, the
agent has no problem finding a path to the goal. Figure \ref{figMaze}b
already displays a more interesting example. The agent has reached
goal {\sf A} and now seeks for goal {\sf B}. However, we blocked the
trespass 1.  Starting at {\sf A} the agent moves normally until it
reaches the blockade.  It stays there and moves slowly up an down in
front of the blockade for a while---this while is of the order of the
low-pass filter time scale $\tau_\k$. During this time, the lateral
weights of the connections pointing to the left are depressed and
after about 150 time steps, this change of weights has enough
influence on the value field dynamics (\ref{valuefield}) to let the
agent chose the way around the bottom to goal {\sf B}.  Figure
\ref{figMaze}c displays the next scene: Starting at {\sf B}, the agent
tries to reach goal {\sf C} again via the blockade 1 (the previous
adaptation depressed only the connections from right to left). Again,
it reaches the blockade, stays there for a while, and then takes the
way around to goal {\sf C}. Figures \ref{figMaze}d and \ref{figMaze}e
repeat this experiment with blockade 2. Starting at {\sf D}, the agent
reaches the blockade 2 and eventually chooses the way around to goal
{\sf E}. Then, seeking for goal {\sf F}, the agent reaches the
blockade first from the left, thereafter from the bottom, then from
the right, then it tries from the bottom again, and finally learned
that none of these paths are valid anymore and chooses the way all
around to goal {\sf F}.  Figures \ref{figMaze}f shows that, once the
world model has re-adapted to account for these blockades, the agent
will not forget about them: Here, moving from {\sf G} to {\sf H}, it
does not try to trespass block 2.

The reader is encouraged to also refer to the movies of the
experiments, deposited at { www.marc-toussaint.net/03-cwm/}, which
visualize much better the dynamics of self-organi\-zation, the planning
behavior, the dynamics of the value field, and the world model
readaptation.

\hide{
\begin{table}[t]\center
\begin{tabular}{ccp{6cm}}
\hline
$\eta$ & .2 & tradeoff between pre-activation and feed-forward activation in
step (3) \\
$\d$ & .9 & decay rate for the error in step (5) \\
$\n$ & .1 & vigilance parameter in step (6) \\
$\m$ & 5 & vigilance parameter in step (8) \\
$\tau$ & 2& time scale for the dynamics (\ref{valuefield}) \\
$\g$ & .98 & discount parameter in (\ref{eqDyn})\\
\hline
\end{tabular}
\caption{\label{tabParas}
Our choice of model parameters.}
\end{table}
}

\hide{

As a first demonstration we choose the discrete maze problem as
illustrated in figure \ref{figMaze}. The action layer comprises four
neurons, each of which encodes one of the directions up, down, left,
and right. In this first experiment the action is given from outside
and the four neurons encode this action bit-wise but with value in
$\{-1,1\}$. The system receives stimuli of 16 bits at each position of
the maze as follows: From each direction, up, down, left, and right,
it receives four bits (again with values $\{-1,1\}$). If the
neighboring place is a wall, these bits were randomly chosen when the
maze was created and corresponds to an observable texture of this
wall; if the neighboring place is free, the four corresponding bits
are all $-1$.  Hence, some state stimuli are ambiguous.  E.g., there
are several places in the maze where all four neighboring places are
free (locations 1 to 4). For all of them the system perceives the same
stimulus. Still it is able to disambiguate these states and represent
them with separate neurons as we will see shortly.

To measure the performance of the system we calculate at each time
step the error between the predicted observation, i.e., the receptive
field of the $(p)$ unit, and the actual stimulus $s$.

In the initial phase of exploration we chose actions completely
random. The model would surely allow to implement more sophisticated
methods of directed exploration in favor of faster knowledge gain
(e.g., exploring where there is a lack of experience). We
have not implemented such strategies yet.

Figure \ref{figErr} displays the learning curve for a typical trial.
Errors always occur when new regions of the maze are explored. E.g., the
lower part of the maze is explored between time 400 and 500 where the
errors are, in average, highest. Although at time 500 every world
state has been explored, the world model is still ill-structured
because ambiguities between state observations are not solved yet.
This leads to the errors between time 500 and 1000.

\begin{figure}[t]\center
\caption{\label{figErr}
  The prediction error during explorative learning. The error is the
  Euclidean error between the 16 bit stimulus with bit values
  $\{-1,1\}$ and the receptive field of the $(p^t)$ neuron. The gray
  curve is the running average of the errors over a time interval of
  10.}
\end{figure}

Figure \ref{figGraphs} displays the CWM layers that have been
developed during this explorative phase at different time steps. It is
important to explain how the layout of these graphs was generated.
Neurons do \emph{not} encode any information on global positions in
the maze; the system perceives only local stimuli and neurons encode
only the respective receptive field that corresponds to these local
stimuli. The graph layout in figure \ref{figGraphs} was thus generated
externally (``by hand'') only to allow more interpretable
illustrations. The algorithm that generated the graph layout was very
simple: Whenever a neuron becomes the best matching unit $z$ store
the current position of the agent in the maze and in the sequel draw
this neuron at that corresponding place.
Discussions are simpler when referring to this geometric layout
instead of an arbitrary 2D-layout that would only preserve the world
model's topology.

The first graph in figure \ref{figGraphs} shows the result of the
first 100 time steps of exploration around location 1 in the maze. The
second graph shows that exploration has extended to the right, and
here the first interesting effect occurs. Locations 1 and 2 induce the
same stimuli and thus the neuron that was generated to represent
location 1 now also represents location 2. That's why some causal
connectivities are wrong and corresponding connections have been
depressed and subsequently deleted (the neuron that represents the
location left to location 1 was completely unlinked and thus also
deleted). Exploration goes on and the third graph exhibits the same
ambiguity problem: Now this neuron represents locations 1, 2 and 3 at
the same time. Between time 300 and 400 the error $l$ for this neuron
has sufficiently accumulated to the belief state duplication for this
unit. The system was successful in relinking one of the clones
according to location 1 and the other according to location 2. Till
time 500 new large areas in the bottom and right of the maze are
explored. Again, locations 3 and 4 are ambiguous with locations 1 and
2 and thus the two neurons that have just been generated to solve the
ambiguity between 1 and 2 are now reused for locations 3 and 4.  The
process goes on until at time 1000 enough neurons to solve the
ambiguities were generated. Only one misleading connection is left
over from these ambiguities, which is later depressed and deleted. At
time 1500 the CWM is complete and stable.

\newcommand{\Time}[1]{\parbox{0ex}{~~~#1}}
\hide{
\begin{figure}[t]\center
\begin{tabular}{p{36mm}|@{~~~}p{36mm}}
 \Time{100}\raisebox{3ex}{\includegraphics*[scale=.3]{\mazehome model/model100}}
&\Time{200}\raisebox{3ex}{~~~~\includegraphics*[scale=.3]{\mazehome model/model200}}\\[2ex]
\hline\\
 \Time{300}\raisebox{3ex}{\includegraphics*[scale=.3]{\mazehome model/model300}}
&\Time{400}\raisebox{3ex}{\includegraphics*[scale=.3]{\mazehome model/model400}}\\[2ex]
\hline\\
 \Time{500}\includegraphics*[scale=.3]{\mazehome model/model500}
&\Time{700}\includegraphics*[scale=.3]{\mazehome model/model700}\\[2ex]
\hline\\
 \Time{1000}\includegraphics*[scale=.3]{\mazehome model/model1000}
&\Time{1500}\includegraphics*[scale=.3]{\mazehome model/model1500}
\end{tabular}
\caption{\label{figGraphs}
The CWM layer during exploration and model building. The time is
indicated for each graph.}
\end{figure}
}


\subsection{Goal-oriented behavior planning}

To demonstrate the behavior planning dynamics we discuss a second
experiment based on the world model that has been learned in the first
experiment. We activate the neural dynamics for behavior planning such
that at each time step the system determines the action to be
executed. To generate the goal stimulus $\vec g$ we picked location
5 and calculated the stimulus the system would receive if located at
this location. This goal stimulus is the input for the neural behavior
planning dynamics (\ref{valuefield}). In the experiments we speed up
relaxation of the dynamics by actually executing 20 iterations of
equation (\ref{valuefield}) every time step.

Figure \ref{figGoal} displays the value field $v_i$ over the CWM for
the 20 times steps after we applied the goal stimulus to the system.
The gray shading of the graph clearly displays the gradient toward the
goal state. Two neurons are colored black in the graph: the one
representing the goal state at location 1 with value $v_g=1$ and
we also colored the actual position of the agent black. Time 0-5 shows
the straight goal-oriented behavior. The agent rests at the goal at
time 6. At this time we switched the goal to location 6. Hence, time
6-11 shows how the value field continuously rearranges to form
the gradient towards location 6. The agent first recognizes this
change of goal at time 7, moving from location 5 to location 1. But
the gradient is not yet monotonous such that at time 8 the agent is
moving back to location 5. From time 8 on, the agent moves straight to
the new goal which is reached at time 21.  Again, we switched the goal
back to location 5, the gradient rearranges, and after one hesitation
at time 24, the agent moves straight back to location 5.

\hide{
\begin{figure}[t]\center
\begin{tabular}{cccc}
time & & & \\
\raisebox{5ex}{0-2}
&\includegraphics*[scale=.15]{\mazehome movie/model2000}
&\includegraphics*[scale=.15]{\mazehome movie/model2001}
&\includegraphics*[scale=.15]{\mazehome movie/model2002}\\[2ex]
\raisebox{5ex}{3-5}
&\includegraphics*[scale=.15]{\mazehome movie/model2003}
&\includegraphics*[scale=.15]{\mazehome movie/model2004}
&\includegraphics*[scale=.15]{\mazehome movie/model2005}\\[2ex]
\raisebox{5ex}{6-8}
&\includegraphics*[scale=.15]{\mazehome movie/model2006}
&\includegraphics*[scale=.15]{\mazehome movie/model2007}
&\includegraphics*[scale=.15]{\mazehome movie/model2008}\\[2ex]
\raisebox{5ex}{9-11}
&\includegraphics*[scale=.15]{\mazehome movie/model2009}
&\includegraphics*[scale=.15]{\mazehome movie/model2010}
&\includegraphics*[scale=.15]{\mazehome movie/model2011}\\[2ex]
& & $\cdots$ & \\
\raisebox{5ex}{18-20}
&\includegraphics*[scale=.15]{\mazehome movie/model2018}
&\includegraphics*[scale=.15]{\mazehome movie/model2019}
&\includegraphics*[scale=.15]{\mazehome movie/model2020}\\[2ex]
\raisebox{5ex}{21-23}
&\includegraphics*[scale=.15]{\mazehome movie/model2021}
&\includegraphics*[scale=.15]{\mazehome movie/model2022}
&\includegraphics*[scale=.15]{\mazehome movie/model2023}\\[2ex]
\raisebox{5ex}{24-26}
&\includegraphics*[scale=.15]{\mazehome movie/model2024}
&\includegraphics*[scale=.15]{\mazehome movie/model2025}
&\includegraphics*[scale=.15]{\mazehome movie/model2026}\\[2ex]
& & $\cdots$ &
\end{tabular}
\caption{\label{figGoal}
The value field over the CWM; black means $V_i^t=1$, white
$V_i^t=0$. The agent's current position is also marked black.}
\end{figure}
}

The experiment demonstrates the functionality of the behavior planning
dynamics. The result is hardly a surprise; given a correct world model
as represented by the CWM, the Value Iteration like dynamics will work
properly.

\subsection{Adaptability to world changes}

The last experiment investigates the system's adaptability w.r.t.\ 
small changes in the world. This small change will be to insert a wall
at location 7 in the maze. In the previous experiment, the agent
always chose the same route to move between the two goal locations 5
and 6. This route trespasses location 7. Blocking the route will
require the agent to adapt its world model, realizing that the
transition through location 7 does no longer exist. If the
reorganization of the CWM is successful, the behavior organization
will automatically find another route to move between locations 5 and
6.

Figure \ref{figBlock} displays such a trial. At time 0 the agent just
reached the goal at location 5, the goal is switched to location 6,
and location 7 is blocked. As before the agent moves the old route
straight toward location 7. At time 11 the agent reaches the crossing
at location 3 and wants to move further to the right. The next time
step it recognizes that it has never experienced the observation of
the state to the left of 7 before. It grows a new representation for
this new observation. It stays at this position since the value
gradient has not yet vanished. The next time step it moves back to
location 3. There, it still perceives the value gradient toward
walking to the right---it tries again at time 15 and predicts again
the observation it was used to from the old route. This fails again
and the connections that led to the wrong prediction are finally
depressed and deleted. Thereafter, the value field relaxes to the
alternative route around the bottom. The agent moves straight along
the new route and reaches the goal at location 6 at time 23.

Then we switch the goal back to location 5. Again, it takes a few time
steps till the value gradient points monotonically towards the
new goal. Incidently, the left-over activation towards moving upwards
lets the agent move up along the old route. At time 26, at the
location to the right of 7, it perceives a new stimulus it never
perceived before. As before it grows a new neuron and unlinks the
connections which led to wrong predictions at times 26-29. Thereafter
the agent moves straight along the new route to goal location 5. The
next time goals are switched, the agent will directly choose the new
route.

\hide{
\begin{figure}[t]\center
\begin{tabular}{@{}p{28mm}@{}p{28mm}@{}p{28mm}@{}}
\Time{11}\includegraphics*[scale=.2]{\mazehome block/model2111}&
\Time{12}\includegraphics*[scale=.2]{\mazehome block/model2112}&
\Time{13}\includegraphics*[scale=.2]{\mazehome block/model2113}\\[2ex]
\Time{14}\includegraphics*[scale=.2]{\mazehome block/model2114}&
\Time{15}\includegraphics*[scale=.2]{\mazehome block/model2115}&
\Time{16}\includegraphics*[scale=.2]{\mazehome block/model2116}\\[2ex]
\Time{17}\includegraphics*[scale=.2]{\mazehome block/model2117}&
\Time{23}\includegraphics*[scale=.2]{\mazehome block/model2123}&
\Time{24}\includegraphics*[scale=.2]{\mazehome block/model2124}\\[2ex]
\Time{25}\includegraphics*[scale=.2]{\mazehome block/model2125}&
\Time{26}\includegraphics*[scale=.2]{\mazehome block/model2127}& 
\Time{27}\includegraphics*[scale=.2]{\mazehome block/model2127}\\[2ex]
\Time{28}\includegraphics*[scale=.2]{\mazehome block/model2130}&
\Time{29}\includegraphics*[scale=.2]{\mazehome block/model2129}&
\Time{30}\includegraphics*[scale=.2]{\mazehome block/model2130}\\[2ex]
\Time{31}\includegraphics*[scale=.2]{\mazehome block/model2131}&
\Time{32}\includegraphics*[scale=.2]{\mazehome block/model2132}&
\Time{33}\includegraphics*[scale=.2]{\mazehome block/model2133}
\end{tabular}
\caption{\label{figBlock}
  Adaptation of the CWM to a small change in the world. The time is
  indicated at the bottom left of each graph. The graph layout w.r.t.\ 
  the new created neurons was done by hand.}
\end{figure}
}

\section{Comparison to other approaches}

Model-free and model-based approaches to Reinforcement Learning have
principally different properties are are thus hardly to compare (see
also \cite{majors-richards:97}). Still, to recall these differences,
let us briefly compare our model to classical, model-free
Reinforcement Learning approaches. The most important point is: What
our model learns is general knowledge about the world which can be
learned without a given fixed goal. After learning, the goal can the
set and changed freely---the system promptly organized its behavior
w.r.t.\ the new goal. In contrast, all what is learned by model-free
Reinforcement Learners (the $V$- or $Q$-function) has only meaning
and in only valid w.r.t.\ one specific goal. If the goal changes
everything has to be relearned. The second point is analogous: For our
model, small changes of the world need only small adaptation of the
CWM. This adaptation automatically changes all the behaviors that
would otherwise conflict with this change of the world. In contrast, a
small change of the world (e.g., a shut door) might require extensive
readaptation of the $V$- or $Q$-function. Since in classical
Reinforcement Learning, these functions are only updated in the course
of exploration, all the possible paths that lead through a shut door
and have been goal-relevant before have to be re-explored in order to
readapt the system to the world change---the system cannot ``grasp''
that the door is shut and that all paths through that door are now
obsolete. Hence, classical Reinforcement Learners would completely
fail on the tasks we used for demonstration: the dynamically changing
goal and the quick adaptation to world changes.

These problems of classical Reinforcement Learners are reduced but not
solved in approaches that fuse model-free Reinforcement Learning with
model-based learning, e.g.\ the Dyna architecture
\cite{sutton:90,sutton:91}. However, also in these approaches Value
Iteration is not executed in parallel but only on the experienced
path. World or goal changes hence also require extensive readaptation
in the course of exploration.

Probably the most similar existing model-based approach is that of
Zimmer \cite{zimmer:96}. A Growing Neural Gas \cite{fritzke:95} is used to
learn a state space representation and here is the first major
difference: action or the action-dependence of state transitions do
not enter the self-organization of the state space model. E.g.\ 
mechanisms like the duplication of belief state representations would
not be possible. On top of the Growing Neural Gas actions are
associated to each connection of the growing neural gas via some extra
associative memory and a ``Reinforcement Learning Manager''. The
adaptation mechanisms are different. A priori knowledge about the
motor kinematics is entangled with the mechanisms in order to fuse
conventional odometry methods with the self-organizing state space
model. (In contrast, our intention was, among others, to learn a this
motor model.)

The work of Shatkay \& Kaelbling
\cite{shatkay-kaelbling:97,shatkay-kaelbling:02} is not concerned
with self-organizing connectionist state space models but presents
elementary work on how to learn action-dependent transition
probabilities. As it is typical in this area, adaptation mechanisms
are based on the EM-algorithms. Particularly interesting is the E-step
where, via the forward-backward algorithm, it is inferred
retrospectively which would have been the better predictions of the
world model. In the M-step, parameters are adapted such that, next
time, these better predictions will be made. Actually, our heuristic
that punishes the \emph{pre}-synaptic unit when a map-prediction
occurred goes a tiny step in the direction of the backward algorithm.
Future work will try to develop the backward algorithm rigorously on
the CWM to allow for a more ``principled'' way of adaptation than the
heuristics given so far. Finally, Thrun
\cite{thrun-moeller-linden:91,thrun:93,thrun:02} uses artificial
neural networks to represent and learn the action-dependent transition
probability.

}

\section{Discussion}

The model we proposed is a connectionist architecture that can
represent and learn the relation between motor signals and perception.
The model is continuous in action, perception, and time domain, does
not presume any a priori knowledge about the motor system, and a
dynamical value field on the learned world model organizes behavior
planning---a method in principle borrowed from classical Value
Iteration. A major feature of our model is its adaptability. The state
space model is developed in a self-organizing way and small world
changes require only little re-adaptation of the CWM. Generally
speaking, the model is a highly functional system based on only local
mechanisms. It demonstrates what these mechanisms can accomplish when
embedded in a suitable structure, e.g., the concrete functionality of
a temporal Hebb rule w.r.t.\ behavior planning in our model, or the
functionality of synapse modulation as we employed it.
\hide{ We believe that such functional understanding of neural
  mechanisms may eventually lead to a highly desirable bridge between
  Machine Learning and Neurosciences.  }

\hide{
All the dynamics and all adaptation rules rely on local interactions
only. Still, the system demonstrates sophisticated functionality that
is of the same order as that of classical approaches to Machine
Learning. It demonstrates all the advantages of model-based approaches
over model-free Reinforcement Learning approaches: General,
goal-independent knowledge about the world is learned; the goal can be
set freely; small, local world changes require only small, local
readaptation of the CWM. The core of the CWM's functional
adaptability lies within its structure, or, more abstract, in the way
its parameters determine behavior. A comparison to a Q-learning
classical Neural Network makes this more clear. Despite sophisticated
back-propagation learning, the way the network weights determine
behavior leads to problems: adaptation is non-local, leading to
cross-talk and catastrophic forgetting
\cite{toussaint:02-ijcnn-theory}; a change of the goal or of the world
requires extensive readaptation of the Q-function.
}

Future work will include the more rigorous probabilistic
interpretations of CWMs which we already indicated in section
\ref{secModel}. Another, rather straight-forward extension will be to
replace random-walk exploration by more directed, information seeking
exploration methods as they have already been developed for classical
world models \hide{schmidhuber:91b,}
\cite{schmidhuber:91c,storck-hochreiter-schmidhuber:95,meuleau-bourgine:98}.
A deeper and still open question is the matter of distributed
representations: Can a procedure analogous to Value Iteration be
generalized to multi-modal representations or representations that are
composed of two layers, each corresponding to another ``perceptual
dimension''? The binding problem is touched here and it seems that the
value dynamics will have to be sequential rather than parallel.

\hide{
The model we proposed is a connectionist architecture that can
represent and learn the relation between motor signals and perception.
In other terms, it learns a world model for partially observable
Markov decision processes. The system does not presume any a priori
knowledge about the motor system. The learned model can be used to
organize behavior planning by means of the dynamical value field---a
method in principle borrowed from classical Value Iteration. A major
feature of our model is its adaptability. The state space model is
developed in a self-organizing way; simple adaptation mechanisms
similar to Hebbian rules allow the local and quick adaptation to small
world changes. Lateral preactivations, which correspond to the
prediction prior in terms of Hidden Markov Models, allow to solve
stimulus ambiguities and thus solved, together with the belief state
duplication mechanism, the partial observability problem in
our demonstration. This exemplifies that state representations should
be developed not only on the basis of the stimulus but rather in tight
relation to the motor signals, as it is the case in our model.
Eventually, the goal should be the self-organization of state
representations due to their relevance for behavior and not purely
based on sensor stimuli.

Future work will explore probabilistic interpretations of CWM, as
indicated in appendix \ref{secEM}, and try to develop a bridge to
Expectation Maximization techniques. More directed exploration
\cite{schmidhuber:91b,schmidhuber:91c,storck-hochreiter-schmidhuber:95,meuleau-bourgine:98}
should be realized also for CWMs. An open question is the matter of
distributed representations: How can a procedure like Value Iteration
be generalized to multi-modal state representations or representations
that are composed of two CWM layers, each corresponding to another
``state dimension''? The binding problem is touched here and it seems
that the value dynamics will have to be more sequential than parallel.
Finally, the self-organization of the CWM topology is very appealing
but might not carry arbitrarily far. Pre-structuring the CWM is
necessary and could, in the tradition of our lab, be done by means of
evolutionary adaptation.
}

\hide{
Comments:
\begin{itemize}
\item Probably it's the changes of observations $P(\D x|\bs')$ rather than the
  observations $P(x'|\bs')$ itself that should be predicted $\too$ add
  $\D x$ to the observations $\iff$ kind of feature extractor of changes!

\item $x$ should have a focus on the relevant observations only. I.e.,
  $O(x|s)$ should be depending on the action $a$ (active
  vision). Actually this is included in the general formalism of
  POMDPs---but I don't know...

  Is action selection in favor of ``seeking for information''
  instead of ``seeking for reward'' also implicit in the Value
  Iteration for planning???? Otherwise: introduce an intermediate
  mechanisms with the goal to seek information and make $\bs$ as
  deterministic as possible (minimize $H(\bs)$ (not as a learning rule
  but as a rule of behavior!))

\item Distributed representations! are still missing.
  
  This is not a problem for the neural structure, because
  $P(\bs'|\bs,a)$ can be represented also for a distributed
  representation of $\bs$. But it is a problem for the decision
  dynamics, the value iteration!!!
  
  The problem would be solved if $Q$ is decomposable: $Q(s,a) =
  Q(s_1,a) + Q(s_2,a)$, $s=(s_1,s_2)$.
  
  ...
  
\item compare to Steinhages competition matrix between different
  elemental behaviors. This implies a distributed representation for
  $a$!!!

\end{itemize}
}

\small
\bibliography{blabla/tex/bibs}
\end{document}